\documentstyle[aaspp4]{article}

\def\sgra	{Sgr~A*}
\def\sgrab	{Sgr~A*~}
\def\tnot	{\ifmmode {\Theta_0}\else {$\Theta_0$} \fi}
\def\rnot	{\ifmmode {R_0} \else {$R_0$} \fi}
\def\onot	{\ifmmode {\Omega_0}\else {$\Omega_0$} \fi}
\def\vsun	{\ifmmode {V_\odot} \else {$V_\odot$} \fi}
\def\peryr	{y$^{-1}$}
\def\perkpc	{kpc$^{-1}$}
\def\porm	{\ifmmode {~\pm~} \else {$~\pm~$} \fi}
\def\kms	{\ifmmode {{\rm km s}^{-1}} \else {km s$^{-1}$} \fi}
\def\msun	{\ifmmode {{\rm M}_\odot} \else  {${\rm M}_\odot$} \fi}
\def\lsun	{\ifmmode {{\rm L}_\odot} \else  {${\rm L}_\odot$} \fi}
\def\vmax	{\ifmmode {V_{max}} \else {$V_{max}$} \fi}
\def\ie		{i.e.,~}
\def\eg		{eg,~}
\def\etal	{et al.~}
\newbox\grsign \setbox\grsign=\hbox{$>$} \newdimen\grdimen \grdimen=\ht\grsign
\newbox\laxbox \newbox\gaxbox
\setbox\gaxbox=\hbox{\raise.5ex\hbox{$>$}\llap
     {\lower.5ex\hbox{$\sim$}}}\ht1=\grdimen\dp1=0pt
\setbox\laxbox=\hbox{\raise.5ex\hbox{$<$}\llap
     {\lower.5ex\hbox{$\sim$}}}\ht2=\grdimen\dp2=0pt
\def\gax{\mathrel{\copy\gaxbox}}
\def\lax{\mathrel{\copy\laxbox}}

\begin{document}

\title{The Proper Motion of \sgra: I. First VLBA Results}
\author{M.~J.~Reid}
\affil{Harvard--Smithsonian Center for Astrophysics, Cambridge, MA 02138}
\authoremail{mreid@cfa.harvard.edu}

\author{A.~C.~S.~Readhead \& R.~C.~Vermeulen}
\affil{California Institute of Technology, Pasadena, CA 91109}
\authoremail{}

\author{R.~N.~Treuhaft}
\affil{Jet Propulsion Laboratory, Pasadena, CA 91109}
\authoremail{}

\begin{abstract}
	We observed \sgrab and two extragalactic radio sources nearby
in angle with the VLBA over a period of two years and measured 
relative positions with an accuracy approaching 0.1 mas.
The apparent proper motion of \sgrab relative to J1745--283 is
$5.90\pm0.4$~mas~\peryr, almost entirely in the plane of the Galaxy.  
The effects of the orbit of the Sun around the Galactic Center can
account for this motion, and any residual proper motion of \sgra, 
with respect to extragalactic sources, is less than about 20 \kms.  
Assuming that \sgrab is at rest at the center of the Galaxy, we estimate that
the circular rotation speed in the Galaxy at the position of the Sun, 
\tnot, is $219\porm20$~{\hbox \kms,} scaled by $\rnot/8.0$~kpc.

	Current observations are consistent with \sgrab containing all of the 
nearly $2.6\times10^6~\msun$, deduced from stellar proper motions, 
in the form of a massive black hole.  	
While the low luminosity of \sgra, for example, might possibly have come
from a contact binary containing of order 10 \msun, the lack of substantial motion
rules out a ``stellar'' origin for \sgra.  
The very slow speed of \sgrab yields a lower limit to the mass of \sgrab of 
about 1,000 \msun.  Even for this mass, \sgrab appears to be 
radiating at less than $0.1\%$ of its Eddington limit.  

\end{abstract}


\section{Introduction}

	\sgrab is a compact radio source, similar to weak active nuclei 
found in other galaxies.  Since its discovery more than two decades ago
by Balick \& Brown (1974), the possibility that \sgrab is a super-massive 
($\sim10^6~\msun$) black hole has been actively considered.  However, its 
radio luminosity of $\approx10^2~\lsun$ (Serabyn \etal 1997)
and its estimated total luminosity of $\lax10^5~\lsun$ are many orders of 
magnitude below that possible from a $\sim10^6~\msun$ black hole.
Thus, on the basis of its spectral energy distribution
\sgrab could be an unusual contact binary containing 
$\sim10$~\msun and radiating near its Eddington limit.  

	Recently, Eckart \& Genzel (1997) and Ghez \etal (1999) 
measured proper motions of stars
near the position of \sgra, as determined by Menten \etal (1997).  
Stellar speeds in excess of 1000 \kms\ at a distance of $0.015$~pc 
from \sgrab indicate a central mass of $2.6\times10^6~\msun$.
While these dramatic results are consistent with the theory that
\sgrab is a super-massive black hole, it is still conceivable that most of 
the central mass could come from a combination of stars and 
perhaps some form of dark matter.  Clearly, independent constraints
on the mass of \sgrab are needed to establish whether or not
it is a super-massive black hole nearly at rest at the dynamical 
center of the Galaxy.

	The {\it apparent} motion of \sgrab can be used to estimate the 
mass and elucidate the nature of this unusual source.
An apparent motion of \sgrab can be attributed to
at least three possible components: 
1) a secular motion induced by the orbit of the Sun about the 
Galactic Center, 
2) a yearly oscillation owing to the Earth's orbital motion around the 
Sun (trigonometric parallax), and
3) a possible motion of \sgrab with respect to the dynamical center of 
the Galaxy.
Measurement of, or limits for, these components of \sgra's apparent motion
can provide unique information on
the circular rotation speed (\tnot) of the Local Standard of Rest (LSR) 
and the peculiar motion of the Sun (\vsun), 
the distance to the center of the Galaxy (\rnot), and 
the nature of \sgrab itself.

	In 1991 we started a program with the Very Long Baseline Array (VLBA)
of the National Radio Astronomy Observatory 
\footnote  {The National Radio Astronomy Observatory is operated by
Associated Universities Inc., under cooperative agreement with the National
Science Foundation.}
(NRAO) to measure the apparent motion of \sgrab .  In principle,
VLBA observations of \sgrab, phase-referenced to extragalactic radio sources, can
achieve an accuracy sufficient to detect secular motions of $\lax1~\kms$ for
a source at the distance of the Galactic Center.
However, achieving this accuracy is quite challenging technically as it involves 
observing at short wavelengths (7 mm), in order to minimize the effects
of interstellar scattering toward \sgra, phase-referencing to extragalactic
sources, and careful modeling of atmospheric effects (because of the
low source elevations).  

	In the early years of the project, we searched for strong, compact, 
extragalactic sources nearby in angle to \sgra and worked toward an
optimum observing strategy.  In this paper, we report results of 
observations spanning a two year period from 1995 to 1997.  Our VLBA images 
clearly show movement of \sgrab with respect 
to extragalactic sources over many synthesized beams.   
While the current positional accuracy is inadequate to determine a 
trigonometric parallax, the secular motion of \sgrab is easily measured.  
This yields an accurate estimate of the angular rotation rate of the Sun
around the Galactic Center, $(\tnot+\vsun)/\rnot$, 
and places interesting limits on the mass of the
black hole candidate responsible for the radio emission.
Our results, and those of Backer and Sramek (1999) from 
observations with the Very Large Array (VLA) over a 15 year period,
indicate that the apparent proper motion of \sgrab is dominated by the 
orbit of the Sun about the Galactic Center and that any peculiar
motion of \sgrab is very small.

\section{Observations}

	Our successful observations using the VLBA were conducted
in the late-night, early-morning periods of 1995 March 4, 1996 March 20 and 31, 
and 1997 March 16 and 27.  (Observations attempted during two August evenings
in 1996 experienced high water vapor turbulence in the atmosphere, and phase
referencing was not successful.)  The observing frequency was 
43.2~GHz and we observed four 8 MHz bands, each at right and left circular 
polarization.  We employed 2-bit sampling at the Nyquist rate, which 
required the maximum aggregate sampling rate supported by the VLBA of 256 
megabits per second.  Only the inner five VLBA stations 
(Fort Davis, Los Alamos, Pie Town, Kitt Peak, and Owens Valley) were used,
as baselines longer than about 1500 km heavily resolve the scatter
broadened image of \sgrab at 43 GHz (\eg Bower \& Backer 1998).

	The observing sequence involved rapid switching between compact 
extragalactic sources and \sgra.
Two sources, J1745--283 and J1748--291, from the catalog of Zoonematkermani et al. 
(1990) were found to be strong enough ($>10$~mJy at 43 GHz) to serve as
reference background sources.  These sources are two of the three used by
Backer \& Sramek (1999) in their program to measure the proper motion
of \sgrab with the VLA; their third background source, J1740--294, proved to
have a steep spectrum and was too weak for inclusion in our 43 GHz program.
We switched among the sources repeating the following pattern:
\sgra, J1745--283, \sgra, J1748--291.
Sources were changed every 15 seconds, typically achieving 7 seconds
of on-source data, except for the earliest observation in 1995 when we were 
experimenting with longer switching times.  We used \sgrab as the 
{\it phase-reference} source, because it is considerably stronger than 
the background sources and could be detected on individual baselines
with signal-to-noise ratios typically between 10 and 20 in the 7 seconds 
of available on-source time.

	We edited and calibrated data using standard tasks in the
Astronomical Image Processing System (AIPS) designed for
VLBA data.  This involved applying data flagging tables generated by
the on-line antenna and correlator systems, station gain curves, and system
temperature measurements.  We solved for station-dependent, intermediate-frequency band
delays and phases on a strong, compact source (NRAO~530).
After applying these corrections, the multi-band data for \sgrab could be combined 
coherently and interferometer phases as a function of time determined.  
These phase solutions were examined by an AIPS task specially written for 
our observations that looked for and flagged data when baseline-dependent phases on 
adjacent \sgrab scans changed by more than one radian.  
Under good weather conditions between 10 and 30\% of the data were 
discarded by this process.  
This provided relatively unambiguous ``phase connection'' for the
remaining data and allowed removal of most of the 
effects of short-term atmospheric fluctuations from all sources.
(We note that during average-to-poor weather conditions, 
our phase measurements on \sgrab every 30 seconds were not frequent 
enough to provide unambiguous phase connection.   Thus, our 15 second switching 
time is probably an optimum trade-off between on-source duty cycle and 
atmospheric coherence losses.)  Data calibrated in this manner produced 
high (\eg 50:1) dynamic range maps of all sources with little or no spatial 
blurring.  The images of the background sources appeared less resolved than
that of \sgra, with no signs of complex or multiple component structures.

	We found that the differences in relative positions between a 
background source and \sgrab for closely spaced ($\approx10$~d) epochs 
were $\approx1$~mas.  These differences exceeded the formal precision, 
estimated by the least-squares fitting process, typically 
by a large factor.  Since the observational conditions and data analysis were nearly
identical for these epochs, small geometric errors (\eg in baselines, source coordinates,
or Earth's orientation parameters) are unlikely to yield position shifts of this 
magnitude given the small angular separation of \sgrab and the background sources.
Therefore, we evaluated the possibility that refractive scattering of the radio
waves in the interstellar medium or modeling errors for the Earth's atmospheric
propagation delay could be responsible.

	Refractive scattering can cause changes in the apparent flux density and
position of a radio source.  Gwinn \etal (1988) published VLBI observations that
limit refractive position wander for water vapor masers in Sgr~B2N, a star
forming region close to the Galactic Center.  For these maser spots, their
22~GHz observations revealed a diffractive scattering size of 0.3~mas and an
upper limit to a Gaussian component of refractive position wander of 0.018~mas.
Theoretical estimates of refractive position wander, based on the diffractive
scattering size, by Romani, Narayan \& Blandford (1986) agree with this limit for
a Kolmogorov electron density spectrum.  Assuming that the refractive effects 
scale as the diffractive scattering size, we expect any refractive wander
of \sgrab at 43~GHz to be $<0.04$~mas, a value about a factor of 25 smaller
than our observed position differences.

	After careful study of the data, 
we concluded that the most likely source of relative position error
is a small error in the atmospheric model used by the VLBA correlator.  
The following simple analysis supports this view:
The phase-delay of the neutral atmosphere, $\tau$, can be approximated 
by $\tau_0 \sec Z$, where $\tau_0$ is the vertical phase-delay and $Z$ is 
the local source zenith angle.
When measuring the {\it difference} in position of two sources separated in 
zenith angle by $\Delta Z$, a first-order Taylor expansion of $\tau$ yields
the expected {\it differenced} phase-delay error for a single antenna:
$$\Delta\tau~\approx~{\partial\tau \over \partial Z} \Delta Z~=~
	\tau_0\sec Z \tan Z \Delta Z~~~~.~~\eqno(1)$$
The seasonally-averaged atmospheric model (Niell 1996) used by the VLBA correlator
is likely to miss-estimate $\tau_0$ by about 0.1 nsec, 
equivalent to a zenith phase-delay of $\approx3$~cm in path length.
This comes mostly from the highly variable contribution by water vapor
(\eg Treuhaft \& Lanyi 1987).
Based on Eq.~(1), this should result in an antenna-dependent error of 
$\Delta\tau \approx 0.3$~cm 
for typical source zenith angles of $\approx70^\circ$ and   
for our typical source separations of $\Delta Z \approx 0.012~{\rm rad} (\approx 0.7^\circ)$.  
Since atmospheric errors are largely uncorrelated for different antennas,
on an interferometer baseline at an observing wavelength of $0.7$~cm, we 
would expect a relative position shift of roughly $\sqrt{2}\times0.3$~cm or
$\approx70$\%\  of a fringe spacing.  This corresponds to 
$\approx0.4$ and $\approx1.7$~mas in the easterly and northerly 
directions, respectively, for our longer baselines.  
This effect should only partially cancel among the different baselines and 
can explain the position errors seen in the raw maps made from observations 
made $\approx10$ days apart.

	In order to improve our relative position measurements, we 
modeled simultaneously our differenced-phase data for the ``J1745--283 minus \sgra'' and
``J1748--291 minus \sgra'' source pairs.  The model allowed for a relative position shift
for each source pair 
and a single vertical atmospheric delay error in the correlator model for
each antenna.   This approach significantly improved the accuracy of 
the relative position measurements as evidenced by the smaller deviations
in relative positions for observations closely spaced in time.  
The vertical atmospheric delay parameters typically indicated a 
correlator model error of a few cm and these parameters were estimated with 
uncertainties of about 1~cm.  Using this approach, we would estimate 
from Eq.~(1) and the above discussion that relative position errors should be 
$\approx0.1$ and $\approx0.4$~mas in the East--West and 
North--South coordinates, respectively, for one day's observation.

	The differenced-phases often displayed post-fit residuals of $\sim 30$ 
degrees of phase, which were correlated over periods of hours.  
Assuming equal and uncorrelated contributions from the
two antennas forming an interferometer pair, this suggests a delay change of about 
0.1 nsec of time or about 3 cm of uncompensated path length.  Since 
typical source zenith angles were about 70 degrees, this corresponds to 
a vertical delay change in the atmosphere above each antenna of about 
0.03 nsec or about 1 cm of path length.
This behavior is consistent with expected large-scale changes in the 
atmospheric delay, and it suggests that significant improvement can
be obtained by monitoring and correcting for  
large scale atmospheric changes.

	The data in Table~1 summarize our relative position measurements.
The data taken on 1995 March 4 were of poor quality, only the stronger of 
the two background sources (J1745--283) was detected, and the positional accuracy 
was significantly worse than for the later epochs.
Positions of the strongest background source, J1745--283, 
phase-referenced to \sgra, for epochs spanning 2 years are 
plotted in Fig.~1 with open circles in the sense \sgrab relative to J1745--283.
They indicate a clear apparent motion for \sgrab relative to J1745--283, consistent in
magnitude and direction with the reflex motion of the Sun around the 
Galactic Center (see \S 3.1).  The positions in the East-West direction have
typical uncertainties of about 0.1 mas, as estimated from the scatter
of the post-fit position residuals about a straight-line motion.  
It is interesting to note that, 
while it takes $\approx220$~My for the Sun to complete an orbit around the 
Galactic Center, the East--West component of the parallax from only 10 days motion 
can be detected with the VLBA!
The position uncertainties in the North-South direction are larger, 
about 0.4~mas, owing to the low declination of the Galactic Center.
	
	The apparent motions of \sgrab relative to J1745--283
over a 2 year time period and J1748--291 relative to J1745--283
over a 1 year time period are given in Table~2.  
The uncertainties in Table 2 include estimates of the systematic
effects, dominated by errors in modeling of atmospheric effects, 
as discussed above.  Assuming that J1745--283 is sufficiently distant that
it has negligible intrinsic angular motion, \sgra's apparent motion is 
$-3.33\pm0.1$ and $-4.94\pm0.4$~mas~\peryr\ 
in the easterly and northerly directions, respectively.
This motion is shown by the solid lines in Fig.~1.

	As a check on the accuracy of our measurements, 
we measured relative positions between the two calibration sources.
These positions are plotted in Fig.~1 (crosses) in the sense 
``J1748--291 minus J1745--283,'' offset to fit the plotting scale
for the ``\sgrab minus J1745--283" data.  
The best fit motions are $+0.17\pm0.14$ 
and $-0.22\pm0.56$~mas~\peryr\ in the easterly and northerly 
directions, respectively, as indicated with the dashed lines in Fig.~1.  
The {\it uncertainty} in the relative motion of J1748--291 with respect to
J1745--283 is $\approx$40\% larger than for 
the motion of \sgrab with respect to J1745--283,because the angular separation 
of the two background sources ($\approx1.0$~deg) is greater than between 
\sgrab and either of the background sources ($\approx0.7$~deg).  
Thus, the background sources display no statistically significant motion 
relative to each other, as expected for extragalactic sources.

	Finally, we have determined the position of \sgrab relative
to an extragalactic source with high accuracy and, therefore, can 
derive an improved absolute position of \sgra.  VLBI observations carried out
by the joint NASA/USNO/NRAO geodetic/astrometric array (Eubanks, private
communication) detected J1745--283 at 8.4 GHz and determined its
position in the U.S.~Navy 1997-1998 reference frame to be
\centerline{J1745--283~~~~~$\alpha$(2000)=17~45~52.4968,~~~$\delta$(2000)=--28~20~26.294~,}
with a uncertainty of about 12~mas.  Assuming this result, we find
the position of \sgrab measured at 1996.25 to be
\centerline{\sgra~~~~~~~~~~$\alpha$(2000)=17~45~40.0409,~~~$\delta$(2000)=--29~00~28.118~.}
The uncertainty in this position is dominated by that of J1745--283.
Note that were one not to correct for the ``large" apparent proper motion of \sgra, 
the position of \sgrab determined for observations made more than 2 years from 
1996.25 would be shifted by an amount greater than the $\approx12$~mas uncertainty.

\section{Discussion}

	The apparent motion of \sgrab with respect to background
radio sources can be used to estimate the rotation of the Galaxy and
any peculiar motion of the super-massive black hole candidate \sgra.
In Fig.~2 we plot the change in apparent position on the plane of the sky of
\sgrab relative to J1745--283.  The dotted line is the variance-weighted 
least-squares fit to the data, and the solid line denotes the orientation
of the Galactic Plane.  Clearly the apparent motion of
\sgrab is almost entirely in the Galactic Plane.
Thus, it is natural to convert the apparent motion from
equatorial to galactic coordinates.  For \sgrab relative to J1745--283,
this yields an apparent motion of
$-5.90\pm0.35$ and $+0.20\pm0.30$~mas~\peryr\ in 
galactic longitude and latitude, respectively.  
The apparent motion in the plane of the Galaxy should be dominated by 
the effects of the orbit of the Sun around the Galactic
Center, while the motion out of the plane should contain only small terms
from the Z-component of the Solar Motion and a possible motion of \sgra.
In the following subsections, we investigate the various components of
the apparent motion of \sgra, place limits on any offset of \sgrab
from the dynamical center of the Galaxy, derive limits on the mass of \sgra,
and constrain the distribution of dark matter in the Galactic Center.

\subsection {Motion of \sgrab in the Plane of the Galaxy}

	Assuming a distance of $8.0\pm0.5$~kpc (Reid 1993), 
the apparent angular motion of \sgrab in the plane of the Galaxy
translates to $-223\pm19$~{\hbox \kms.}  The uncertainty includes the effects of
measurement errors and the 0.5~kpc uncertainty in \rnot.
Provided that the peculiar motion of \sgrab is small (see \S 3.2), 
this corresponds to the reflex of true orbital motion of the Sun around the 
Galactic Center.  This reflex motion can be parameterized
as a combination of a circular orbit (\ie of the LSR)
and the deviation of the Sun from that circular orbit (the Solar Motion).
The Solar Motion, determined from Hipparcos data 
by Dehnen \& Binney (1998), is $5.25\pm0.62$~\kms\ in the 
direction of galactic rotation.
Removing this component of the Solar Motion from the {\it reflex} of 
the apparent 
motion of \sgrab yields an estimate for \tnot\ of $218\pm19$~{\hbox \kms.}  
This value is consistent with most recent estimates of about 220~\kms 
(Kerr \& Lynden-Bell 1986) and can be scaled for different values of the 
distance to the Galactic Center by multiplying by $\rnot/8$~kpc.

	The most straightforward comparison of our direct measurement of the 
{\it angular} rotation rate of the LSR at the Sun (\tnot/\rnot)
can be made with Hipparcos measurements based on motions of Cepheids.
Feast \& Whitelock (1997) 
conclude that the angular velocity of circular rotation at the Sun,
\tnot/\rnot (= Oort's A--B), is $27.19 \porm 0.87$~\kms\ \perkpc\ 
($218 \porm 7$~\kms\ for $\rnot=8.0$~kpc).  
Our value of $\tnot/\rnot$, obtained by removing the Solar Motion in
longitude from the {\it reflex} of the motion of \sgrab in longitude,
is $27.2\pm1.7$~\kms\ \perkpc.
The VLBA and Hipparcos measurements are consistent within their joint errors, 
and both measurements are insensitive to the value of \rnot, as it is only used 
to remove the small contribution of the Solar Motion.  
It is important to note that our value of $\tnot/\rnot$ is a 
true ``global'' measure of the angular rotation rate of the Galaxy.  
The consistency of the local (A--B) and global measures of $\tnot/\rnot$ suggests that 
local variations in Galactic dynamics ($d\tnot/d\rnot$) are less than the joint 
uncertainties of about 2~\kms~\perkpc.

	After removing the best estimate of the motion of the Sun around the
Galactic Center, our VLBA observations yield an estimate of
the peculiar motion of \sgrab of $0.0\pm\sqrt{0.87^2+1.7^2}$~\kms~\perkpc\ or
$0\porm15$~\kms\ towards positive galactic longitude.
This estimate of the ``in plane'' motion of \sgrab comes from differencing
two {\it angular} motions.  Since this difference is negligible, the uncertainty 
in \rnot\ does not affect this component of the peculiar motion of \sgra.
Given the excellent agreement in the global and local measures of the angular 
rotation rate of the Galaxy, and the lack of a detected peculiar motion for \sgra,
it is likely that \sgrab is at the dynamical center of the Galaxy.

\subsection	{Motion of \sgrab out of the Plane of the Galaxy}

	Whereas the orbital motion of the Sun (around the Galactic Center)
complicates estimates of the ``in plane" component of the 
peculiar motion of \sgra, motions out of the plane are simpler to interpret.  
One needs only to subtract the small Z-component of the Solar Motion from the
observed motion of \sgrab to estimate the out-of-plane component of the peculiar motion
of \sgra.   An implicit assumption in this procedure is that the 
Solar Motion reflects the true peculiar motion of the Sun.  Since most estimates of the 
Solar Motion are relative to stars in the solar neighborhood, this
assumes that ``local'' and ``global'' estimates of the Solar Motion 
are similar.   This procedure could be compromised slightly were the 
solar neighborhood to have a significant motion out of the 
plane of the Galaxy, owing, for example, to a galactic bending or corrugation mode.

	One way to limit the magnitude of a possible difference between a local and global
estimate for the Solar Motion is to compare motions based on nearby stars 
with those based on much more distant stars.  Using stars within about 
0.1~kpc Dehnen \& Binney (1998) find the Z-component for the Solar Motion
to be $7.17 \pm 0.38~\kms$, while Feast \& Whitelock (1997) determine a value
of $7.61 \pm 0.64~\kms$ for stars with distances out to a few kpc.
Since these values agree within their joint uncertainties of about 0.74 \kms, 
it seems unlikely that local and globlal values for the Solar Motion could differ by 
more than about 1~\kms.

	The Hipparcos measurements of large numbers of stars in the solar
neighborhood provide an excellent reference for determining the local solar motion.
We adopt the value of Dehnen \& Binney (1988), which comes from the velocities of 
more than 10,000 stars within about 0.1~kpc of the Sun.
Removing $7.17$~\kms from our measured apparent motion of 
\sgrab out of the plane of the Galaxy, we estimate the peculiar motion of 
\sgrab to be $15\pm11$~\kms\ toward the north galactic pole (see Table 3).  
The uncertainty is dominated by our proper motion measurements and can
be greatly improved by future measurements.  Note, for example, 
that increasing the weight (decreasing the estimated uncertainty) of the 
1995 measurement would decrease the magnitude of our peculiar motion estimate.
We do not consider our estimate of the peculiar motion of \sgrab out of the
plane of the Galaxy to be statistically distinguishable from a null result.  

\subsection {Limits on the Mass of \sgra}

	Our estimates of a peculiar motion of \sgrab 
provide an upper limit of about $20$~\kms each 
for motions in and out of the galactic plane.  
Since stars in the inner-most regions of the central
cluster move at speeds in excess of 1000~\kms\ (Eckart \& Genzel 1997, 
Ghez \etal 1999), a central ``dark mass'' of approximately 
$2.6\times10^6$~\msun contained within 0.015 pc of \sgrab seems required.
It is likely, but unproven, that most of this mass is contained in a 
super-massive black hole: \sgra.  
Given the fact that independent measurements (Backer \& Sramek 1999, 
and this paper) show that \sgrab moves at least two orders of magnitude slower
than its surrounding stars, \sgrab must be much more massive than the 
$\sim10~\msun$ stars observed in the central cluster.  
(See also Gould \& Ram\'irez [1998] for discussion of the
implications of a lack of apparent {\it acceleration} of \sgra.) 
In this section we derive a lower limit to the mass of \sgrab and constrain
possible distributions of dark matter, not in the form of a super-massive
black hole.

\subsubsection	{Virial Theorem}

	Unfortunately, the Virial theorem is of little help in relating the
masses and velocities of stars to that of a central massive black hole.
For Virial equilibrium, 
$$T_{s} + T_{bh} = -{1\over2}~(U_{s} + U_{bh})~~~,\eqno(2)$$
where $T$ and $U$ correspond to the kinetic and potential energy terms
and the subscripts $s$ and $bh$ identify those associated 
with the stars and a central, massive, black hole, respectively.  
Essentially all the kinetic 
energy can be tied up in the stars ($T_{s}$) and all the gravitational 
potential energy found associated with the black hole ($U_{bh}$).  
In this case, attempts to estimate the kinetic energy of the 
black hole ($T_{bh}$)
require differencing two large and uncertain quantities and will be 
essentially useless.

\subsubsection	{Equipartition of Kinetic Energy}

	Upper limits on the motion of \sgrab have been used to infer lower
limits on the mass of \sgrab by assuming equipartition of kinetic energy
(\eg Backer 1996, Genzel \etal 1997).
This is reasonable for stellar systems such as globular clusters, where
massive stars are found concentrated toward the cluster center and move
more slowly than lower mass stars.
Similar results might also hold for a system involving a central black hole
and a surrounding stellar cluster, provided the core mass of the cluster
{\it greatly exceeds} that of the black hole.  
However, given the likely mass dominance of \sgrab over the stars 
within 0.015 pc, where high stellar speeds have been measured, 
equipartition of kinetic energy may be an unreliable approximation.  

	Indeed, our solar system may prove a better ``scale model'' (with planets 
corresponding to stars and the Sun corresponding to \sgra).  
The Sun orbits the barycenter of the solar system, approximately
in a binary orbit with Jupiter.  Neglecting small perturbations from
other planets, for a binary orbit in the center of mass frame, momentum conservation
requires that $m_J v_J = M_\odot V_\odot$, where the subscripts $J$ and $\odot$ refer
to Jupiter and the Sun, respectively.  The ratio of the kinetic energy of
Jupiter to the Sun is given by $m_J v_J^2 / M_\odot V_\odot^2 = M_\odot / m_J$.
Hence, the kinetic energy of Jupiter exceeds that of the Sun by a factor
equal to the inverse of the ratio of their masses and 
equipartition of kinetic energy does not apply.

\subsubsection {Case 1: $M_{SgrA*} \sim 2.6\times10^6$~\msun}

	In order to better evaluate how an upper limit to the motion of \sgrab
can be used to provide a lower limit to its mass, we carried out N-body 
simulations of stars orbiting about a massive black hole.  
We used a simple, direct integration code (NBODY0) of Aarseth (1985), 
documented by Binney \& Tremaine (1987) and modified for our purposes.
Initial simulations used 255 stars orbiting a $2.6\times10^6~\msun$ black hole.
The stellar masses were chosen randomly to represent the upper end of a 
stellar mass function with a power law distribution from 20 down to 2 \msun.  
The number of stars and their masses used in this simulation are comparable
to those observed within the central 0.5~arcsec, or 4000~AU (Genzel \etal 1997).
Stellar orbits were chosen by randomly assigning a distance from the black hole,
uniformly distributed in the range 10 to 10,000 AU, calculating a circular 
orbital speed, and then adjusting the speed randomly by between $\porm20\%$ 
of the circular speed for each of the three Cartesian coordinates.  
Before starting the N-body 
integrations, the orbital orientations were randomized by rotating the 
coordinates (and velocity components) through three Euler rotations with 
angles chosen at random.

	The N-body simulations show quasi-random motions of the massive black 
hole.  After relatively short periods of time ($\ll10,000$~years) a 
``steady state'' condition appeared to be reached.  The speed of 
a typical star was about 700 \kms\  at an average distance of 
6,000 AU.  The motion of \sgrab changed completely in all three coordinates 
on time scales $\ll100$~years and was typically $\lax0.1$~\kms\ 
in each coordinate.  The rapid, but bounded, changes in the motion of \sgrab 
suggests that close encounters with individual stars are responsible for most
of the observed motions.  Assuming this is the case, one can make a 
simple analytical estimate of the expected motion of \sgra, owing to close
encounters with stars in the dense central cluster.

	For a two-body interaction conserving momentum and viewed in the center 
of mass frame,
                         $$mv = MV~~~,\eqno(3)$$
where $m$ and $v$ are the mass and speed of the star
and $M$ and $V$ are the mass and speed of the black hole, respectively.
For the case of interest where $M \gg m$, the orbital speed of the star 	
at periastron, $v_p$, is given by the well known relation
   $$v_p^2 = { GM \over a } \Bigl( {1+e \over 1-e} \Bigr)~~~,\eqno(4)$$
where $G$ is the gravitation constant, $a$ is the stellar semi-major axis,
and $e$ is the orbital eccentricity.
Defining $V_p$ as the speed of the black hole at periastron, combining
Eqs.~(3) and (4) yields
$$V_p = \Bigl( {m \over M} \Bigr) \Bigl( {GM (1+e) \over a (1-e)} \Bigr)^{1/2}~~~.
               \eqno(5)$$

	Our observations are only sensitive to orbital periods longer 
than of order 1 year.  Such orbital periods occur for stellar 
semimajor axes greater than about 50 AU (1000 Schwarzschild radii)
for a $2.6\times10^6 \msun$ black hole.
Thus, for our application reasonable values for the parameters in Eq.~(5) are as follows:
$m \sim 10~\msun$, $e \sim 0.5 $, and $a \sim 50$~AU, and the expected
orbital speed of \sgrab would be $\approx0.03$~\kms.  
(We adopt the periastron speed, instead of the lower average orbital speed,
because the influence of many orbiting stars will likely increase the speed
of \sgra, compared to the single star result.)
This speed is
well below our current limit for the motion of \sgra.
Thus, the simplest interpretation consistent with the fast stellar motions
and the slow \sgrab motion is that \sgrab is a super-massive black hole.

\subsubsection {Case 2: $M_{SgrA*} \ll 2.6\times10^6$~\msun}

	While the simplest interpretation is that \sgrab is a 
$2.6\times10^6$~\msun black hole, our upper limit on any peculiar motion 
for \sgrab currently is two to three orders of magnitude above its expected
motion for that mass.   Thus, it seems reasonable to investigate the possibility that
the mass within a radius of 0.01 pc 
is not dominated by 
\sgra, but instead is in some form of ``dark'' matter.  In this
case, \sgrab will react to the gravitational potential and orbit
the center of mass of this dark matter.  Below we show that the upper
limits on the motion of \sgrab are complimentary to the stellar
proper motion results and strongly constrain both the mass of \sgrab 
and any possible configuration of matter within the central 0.01 pc.

	Fig.~3 displays the enclosed mass versus radius for 
four mass models that are consistent with the stellar proper motions
(cf., Genzel \etal 1997, Ghez \etal 1999).
These models all yield flat ``enclosed-mass versus radius" relations at
distances $\gax0.01$~pc from the center of mass of the system where
measurements exist.  The most centrally condensed
mass distribution, a point mass, is shown as the horizontal dash-dot
line labeled ``a''.  The least centrally condensed mass distribution plotted is for
a Plummer density distribution, where density, $\rho$, is given by
$\rho=\rho_0~\bigl(1+(r/r_0)^2\bigr)^{-\alpha/2}$,
for $\rho_0=6\times10^{11}$~\msun pc$^{-3}$, $r_0=0.01$~pc, and $\alpha=5$.  
This distribution is shown with a curved dash-dot line labeled ``d''.
It is difficult to make a physically reasonable mass
distribution that is significantly less centrally condensed than this
and consistent with the stellar motion data.  These two ``extreme" model distributions
approximately bound all allowed mass distributions; two particular
examples of intermediate models are shown in Fig.~3 with dashed curves
labeled ``b'' and ``c''.

	Assuming that \sgrab has a mass $\ll10^6$~\msun, it will
orbit about the center of mass of the system.  Orbital speed for
a body at a radius, r, from the center of mass is given by
$V = \sqrt{{GM_{encl}/r}},$\ 
where $M_{encl}$ is the enclosed 
mass at that radius.  Setting $V=20$~\kms, our limit for the motion
of \sgra, 
produces the sloping solid line in Fig.~3.  
Only enclosed masses below that line are permitted by our observations.  
For radii greater than about $3\times10^{-5}$~pc (6 AU), this limit rules 
out all but the {\it least} centrally condensed mass models.
For radii less than about $3\times10^{-5}$~pc, the orbital motion
of \sgrab produces angular excursions less than 0.8~mas.  In this
case, while the orbital speed might greatly exceed 20~\kms, we may
have failed to detect these excursions owing to our poor 
temporal sampling and $\approx0.4$~mas errors in the North--South direction.
Thus, we do not extend the motion limit line below $3\times10^{-5}$~pc,
and at this radius we replace the motion limit with a vertical line 
in Fig.~3.

	The stellar proper motions and the limits on the proper motion 
of \sgrab combine to exclude almost all of ``parameter space" for 
models of the density distribution of material in the inner
0.1 pc of the Galactic Center.   
The stellar motions exclude ``soft'' gravitational potentials
(\ie the least centrally condensed mass distributions) and the motion
limit for \sgrab excludes ``hard'' gravitational potentials.
Continued VLBA observations of
\sgrab over the next five years could reduce the uncertainty in
the peculiar motion of \sgrab to about 2~\kms (dotted line in Fig.~3)
out of the plane of the Galaxy.  
Improved accuracy for positions in individual VLBA observations, 
necessary for a trigonometric parallax of \sgra, could move the
small angular excursion limit to $<0.2$~mas.  This would further and
drastically restrict the range of possible models for a dominant central
dark matter condensation.  

	The point-like mass distribution labeled ``a" in Fig.~3 essentially 
requires a super-massive black hole.  Since we have assumed in this section 
that this is {\bf not} \sgra, we would be left with the question of
why a low-mass \sgrab radiates far more than a super-massive black hole in
essentially the same environment.
Other mass models, and especially those with the most centrally condensed 
mass distributions (\eg labeled ``b" in Fig.~3) require exceedingly high
mass densities.  For example, model ``b'' has $\approx5\times10^5$~\msun\ 
within a radius of 6~AU, resulting in a density of $10^{19}$~\msun~pc$^{-3}$.
Theoretical arguments suggest that such models are unlikely to be 
stable for even $10^7$~y, regardless of the composition of the matter (Maoz 1998).  

	The hardest cases to exclude on either observational or theoretical
grounds are given by the mass distributions that are the least centrally condensed,
but still consistent with the stellar motion data.  For this case, one can ask 
the question, at what mass will perturbations by stars in the central cluster 
lead to detectable motion of \sgra?  
In order to answer this question, we modified the N-body code described in
\S~3.3.3 to include
a fixed gravitational potential appropriate for a Plummer law mass distribution
with  $\alpha=5$, $\rho_0=6\times10^{11}$~\msun~pc$^{-3}$, and $r_0=0.01$~pc
(model ``d").
\sgrab was assigned a mass ($\ll10^6$~\msun) and the entire system,
including 254 stars was allowed to evolve in time.  
For the softer allowable gravitational potentials (\eg ``c'' and ``d'')
there is little enclosed mass within $\sim10^{-4}$~pc to bind \sgra.
We found that when \sgra's mass was less than $\approx3,000$~\msun, \sgrab gradually
moved outward from the center of the gravitational potential and achieved
orbital speeds in excess of 20~\kms.  Thus, we conclude that the
lack of detectable motion for \sgrab places a conservative lower limit of 
about 1,000~\msun for the mass associated with \sgra.

\section	{Conclusions}

	The {\it apparent} proper motion of \sgra, relative to extragalactic sources,
is consistent with that expected from the Sun orbiting the center of the Galaxy.
Thus, \sgrab must be very close to, and most likely at, the dynamical center of the 
Galaxy.  In this case, the proper motion measurement gives the angular rotation 
speed at the Sun, $(\tnot+\vsun)/\rnot$, directly, from which we estimate
$\tnot=218\pm19$~\kms for $\rnot=8$~kpc.

	Our lower limit for the peculiar motion of \sgrab of about 20~\kms implies
a lower limit for mass of \sgrab of $\sim10^3$~\msun.  This rules out the
possibility that \sgrab is any known multiple star system, such as a contact
binary containing $\sim10$~\msun\ and radiating near its Eddington luminosity.
A mass of more than $\sim10^3~\msun$ and a luminosity 
$\lax10^5~\lsun$ indicates that \sgrab is radiating at
$\lax0.1\%$ of its Eddington limit.  

	All observations are consistent with 
\sgrab being a super-massive black hole.  Since the lower limit for
the mass of \sgrab is only about 0.1\% of the gravitational mass inferred from the
stellar motions, one cannot claim from our observations alone
that even a significant fraction of the dark mass must be in \sgra.  
However, alternative models involving ``dark'' matter distributions are
severely restricted by observations.

	Future VLBA observations should be able to reduce the uncertainty in 
the measurement of the motion of \sgrab out of the Galactic plane 
to $\sim0.2$~\kms, at which point 
knowledge of the Solar Motion may become the limiting factor.  
Should the peculiar motion of \sgrab be less than 0.2 \kms, 
then its mass almost certainly exceeds $\sim10^5~\msun$.  
Such a large mass tied {\it directly} to the
radio source, whose size is $\lax1$~AU from VLBI observations,
would be compelling evidence that \sgrab is a super-massive black hole.

\acknowledgments
We thank D. Backer for providing coordinates for sources J1745--283 and J1748--291
early in our project, M. Eubanks for measuring the astrometric position of J1745--283,
and V. Dhawan for helping with the VLBA setup.


\begin{deluxetable}{llccl}
\tablenum{1}
\tablewidth{0pt}
\tablecaption{Residual Position Offsets Relative to \sgra}
\tablehead{
\colhead{Source}         & \colhead{Date of}      &
\colhead{East Offset$^a$}& \colhead{North Offset$^a$}&
 \\
\colhead{}               & \colhead{Observation}   &
\colhead{(mas)}       & \colhead{(mas)    }  &
 }
\startdata
J1745--283& 1995 March \phantom{0}4	& \phantom{0}0.73& \phantom{0}1.18\nl
                                                        \nl
         & 1996 March 20& \phantom{0}3.60& \phantom{0}5.80\nl
         & 1996 March 31& \phantom{0}3.75& \phantom{0}6.64\nl
                                                        \nl
         & 1997 March 16& \phantom{0}6.90& 11.18\nl
         & 1997 March 27& \phantom{0}7.10& 11.13\nl
                                                        \nl
J1748--291& 1996 March 20& 27.55& 82.07\nl
         & 1996 March 31& 27.53& 81.98\nl
                                                        \nl
         & 1997 March 16& 31.00& 86.92\nl
         & 1997 March 27& 31.09& 86.60\nl
                                                        \nl
\tablenotetext{a}{Uncertainties are 0.1 and 0.4~mas for the
East and North offsets, respectively, for all epochs except 1995 March 4
which are estimated to be 0.5 and 0.8~mas owing to poor phase coherence.}
\tablecomments{Positions are relative to \sgra, after removing the
$\approx$0.7 degree differences.  The origins are based on our originally
adopted J2000 positions for \sgrab (17 45 40.0500, --29 00 28.120),
J1745--283 (17 45 52.5056, --28 20 26.302), and J1748--291 
(17 48 45.6930, --29 07 39.488).}
\enddata
\end{deluxetable}


\begin{deluxetable}{lllllll}
\tablenum{2}
\tablewidth{0pt}
\tablecaption{Apparent Relative Motions}
\tablehead{
\colhead{Source~--~Reference}         & 
\colhead{East Offset}    & \colhead{North Offset} &
\colhead{$\ell^{II}$ Offset}    & \colhead{$b^{II}$ Offset} &
 \\
\colhead{}               & 
\colhead{(mas y$^{-1}$)}       & \colhead{(mas y$^{-1}$)}  &
\colhead{(mas y$^{-1}$)}       & \colhead{(mas y$^{-1}$)}  &
 }
\startdata
\sgra~~~~~--~J1745--283 &$-$3.33~(0.10) &$-$4.94~(0.40) 
			&$-$5.90~(0.35) &$+$0.20~(0.30)\nl
                                           \nl
J1748--291~--~J1745--283&$+$0.17~(0.14) &$-$0.22~(0.56) 
			&$-$0.10~(0.56) &$-$0.26~(0.42)\nl

\tablecomments{Motions are based on data in Table 1, and
estimated uncertainties are given in parentheses.
}

\enddata
\end{deluxetable}


\begin{deluxetable}{lllll}
\tablenum{3}
\tablewidth{0pt}
\tablecaption{\sgra's Motion in Galactic Coordinates$^a$}
\tablehead{
\colhead{Description}    & 
\colhead{$\ell^{II}$}    & \colhead{$b^{II}$} &
 \\
\colhead{}               & 
\colhead{(km s$^{-1}$)}       & \colhead{(km s$^{-1}$)}  &
 }
\startdata
Observed \sgra\  motion			&--223~(19)	&\phantom{1}8~(11) \nl

Effects of Solar Motion$^b$ removed		&--218~(19)	&15~(11) \nl

Effects of Galactic Rotation$^c$ removed	&\phantom{--21}0~(15)	&15~(11) \nl

\tablenotetext{a}{Motions are for \sgra, based on the J1745--283 (relative to \sgra) 
data in Table 2.  Speeds assume $\rnot=8.0\pm0.5$ kpc (Reid 1993).  
Quoted uncertainties given in parenthesis are 1-$\sigma$ and include measurement
uncertainty and an angular-to-linear motion scaling error
from the uncertainty in \rnot.}

\tablenotetext{b}{Adopted Solar Motion with respect to a circular orbit
is $5.25\pm0.62$ \kms\ in $\ell^{II}$ and $7.17\pm0.38$ \kms\ 
in $b^{II}$ (Dehnen \& Binney 1998).}

\tablenotetext{c}{Adopted circular rotation of
$27.19\pm0.87$ \kms\ \perkpc\ (Feast \& Whitelock 1997) removed from
our measured angular rotation rate of $-27.2\pm1.9$~\kms~\perkpc\ (after
correction for the Solar Motion) and then multiplied by $\rnot=8.0$~kpc.  
Since the angular motion difference is small, it is not sensitive 
to the value adopted for \rnot\ when converting to linear motion.
The quoted uncertainty is dominated by measurement uncertainties for the angular motions,
scaled by \rnot.}

\enddata
\end{deluxetable}

\vfill\eject

\centerline	{~~~~~}

\includegraphics{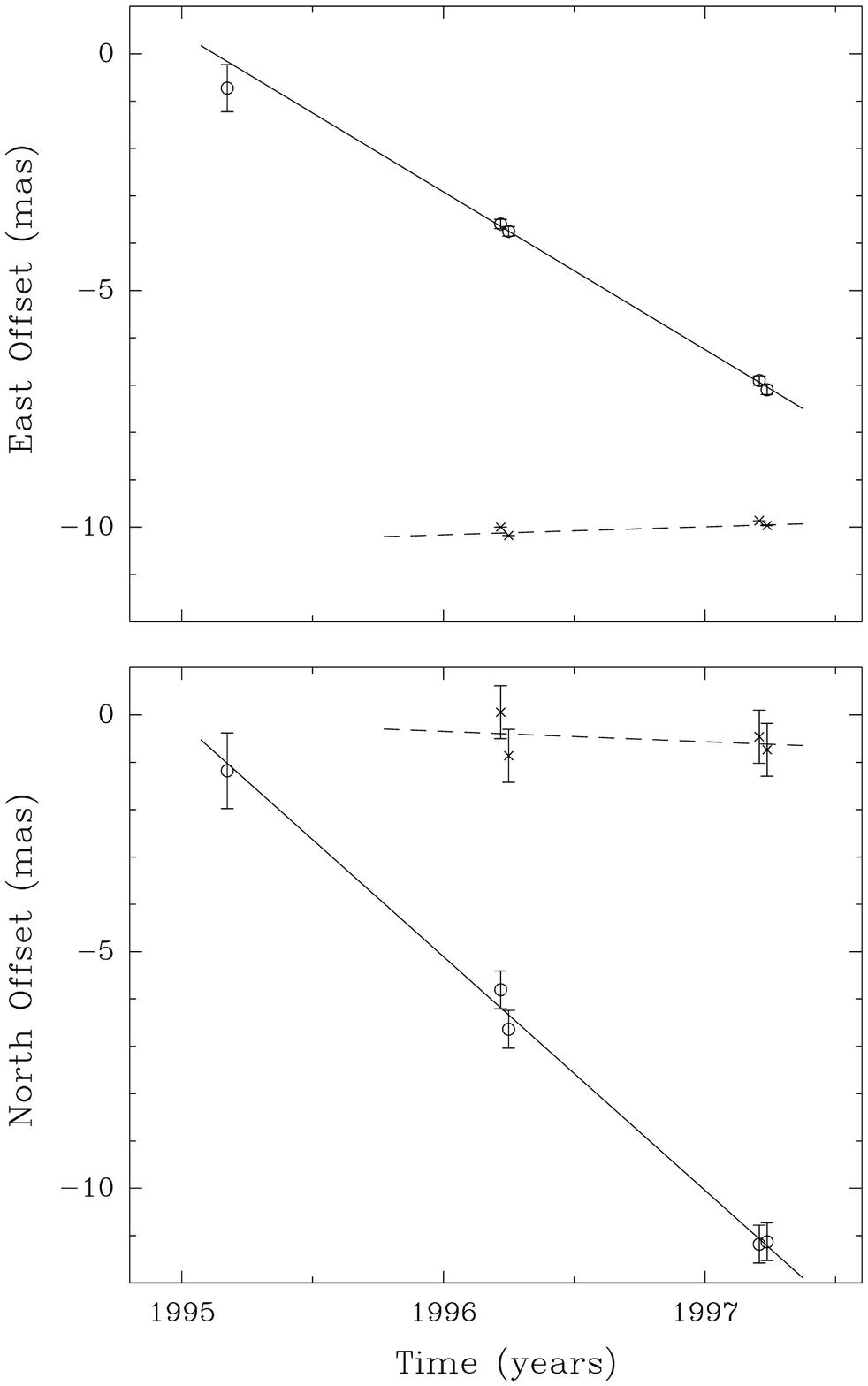} 

\vskip 7truein
\par
~~~~~~~~Fig.~1.~~Position residuals of \sgrab\ relative to J1745--283 (circles) and
J1748--291 relative to J1745--283 (crosses) versus time.  Eastward
components are shown in the top panel and Northward components in the
bottom panel.  The solid and dashed lines give the variance-weighted
best fit components of proper motion.  The J1748--291---J1745--283
positions have been offset to fit the plot scale for the
\sgra---J1745--283 data.
\vfill\eject

\centerline	{~~~~~}
\includegraphics{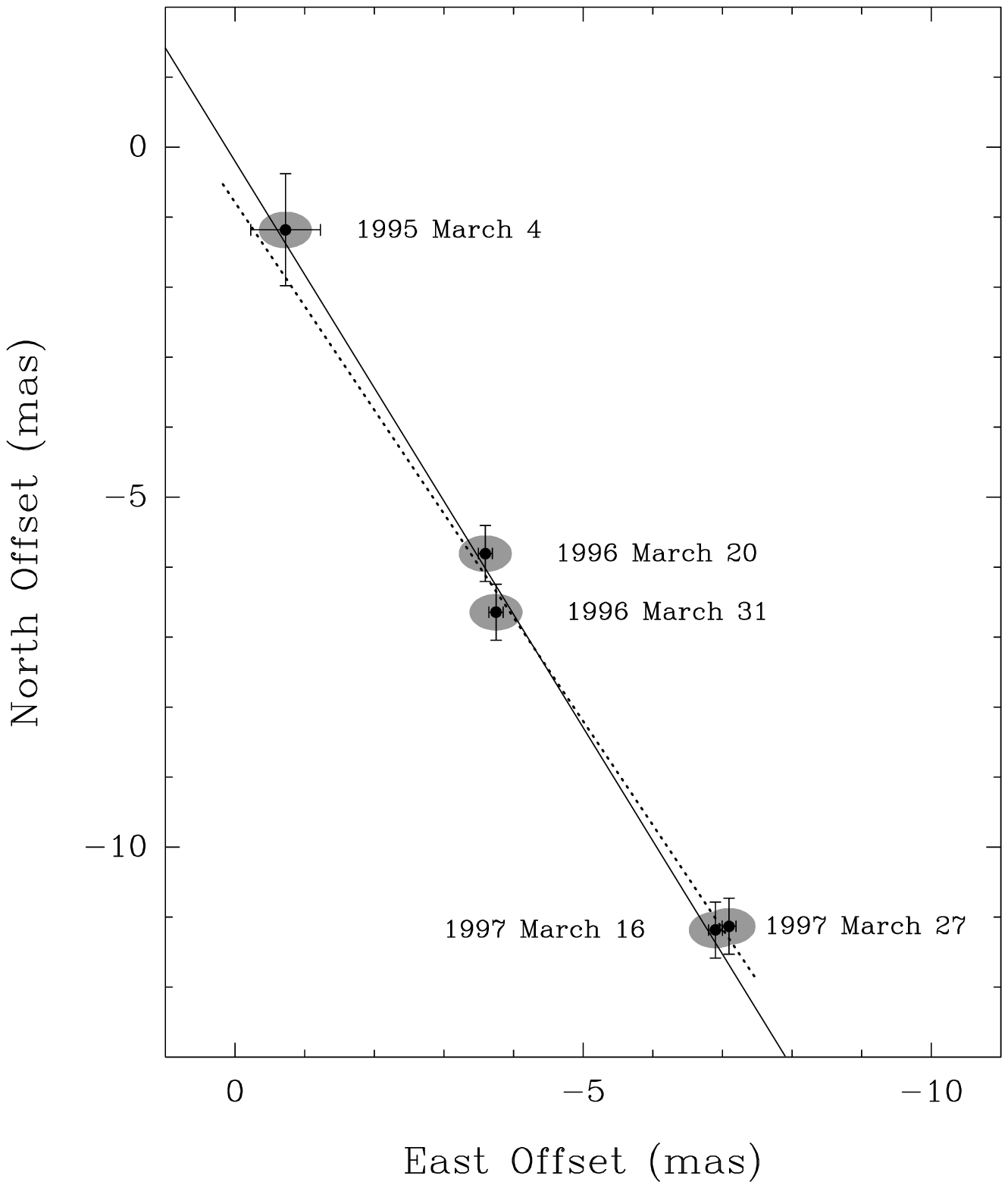} 
\vskip 6.2truein
\par
~~~~~~~~Fig.~2.~~Position residuals of \sgrab\ relative to J1745--283 on the
plane of the sky.  North is to the top and East to the left.  Each
measurement is indicated with an ellipse, approximating the
apparent, scatter broadened size of \sgrab at 43 GHz, the
date of observation, and $1-\sigma$ error bars.
The dashed line is the variance-weighted best-fit proper motion,
and the solid line gives the orientation of the Galactic plane.
\vfill\eject

\centerline	{~~~~~}
\includegraphics{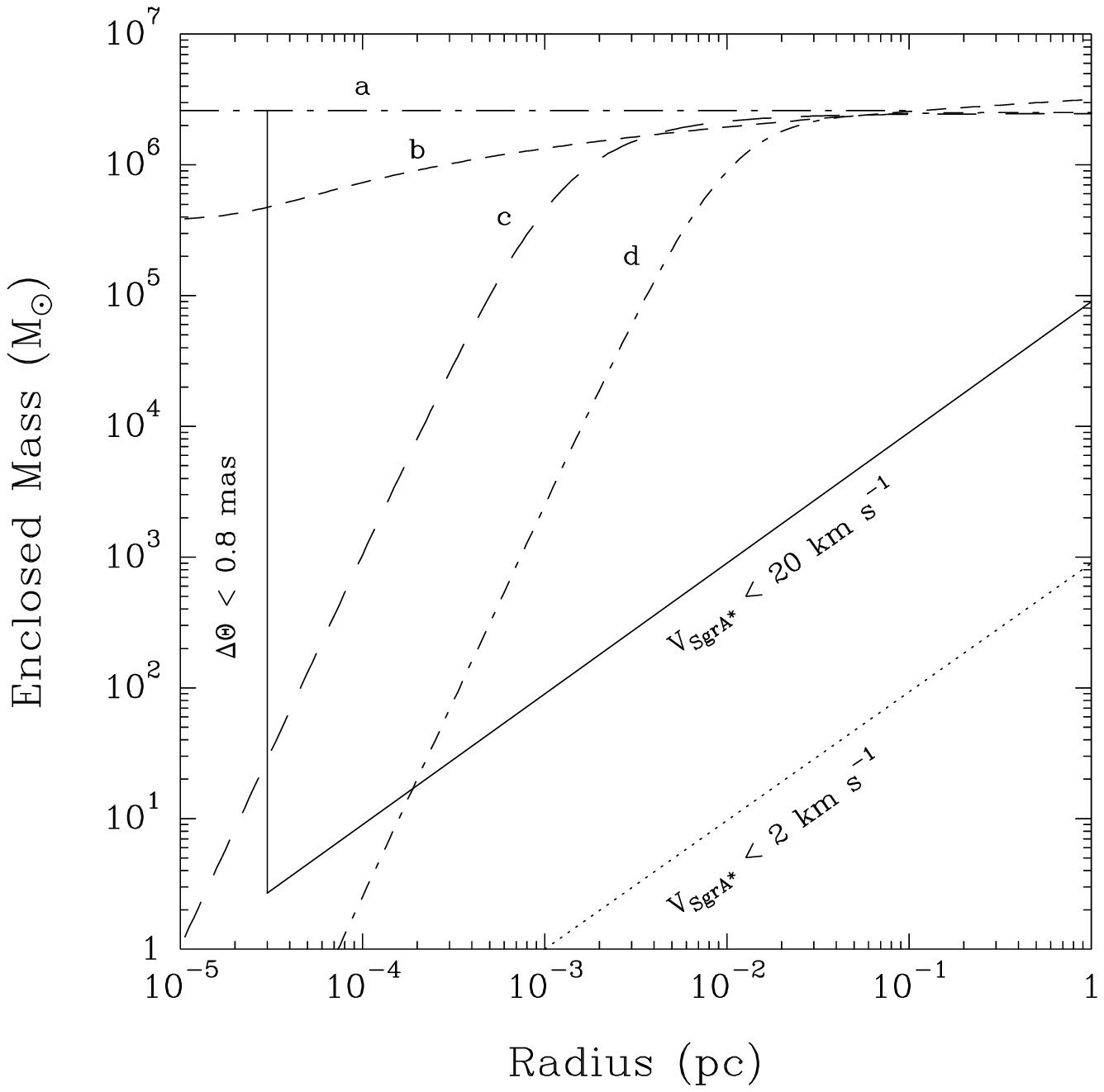} 
\vskip 5.5truein
\par
~~~~~~~~Fig.~3.~~Enclosed mass versus radius for various model distributions
of dark matter, {\it assuming the mass of} \sgrab $\ll 10^6$~\msun.
Models labelled ``a'' through ``d'' have decreasing central mass condensations
(progressively softer gravitational potentials) and approximately bound
mass distributions that are consistent with stellar proper motion data.
Model ``a'' is a point mass; model ``b'' through ``d'' have Plummer density
distributions with $\rho_0$ of $3.9\times10^{18}, 2.5\times10^{14}, {\rm and}~6.0\times10^{11}$
\msun~pc$^{-3}$; $r_0$ of 0.00002, 0.001, and 0.01 pc; and $\alpha$ of 3, 4, and 5,
respectively.  The sloping solid line indicates the upper limit for enclosed mass based
on the proper motion of \sgra.  The vertical solid line at $3\times10^{-5}$~pc (6~AU)
indicates the upper limit in radius, where angular excursions of \sgrab of 
$<0.8$~mas 
could be missed owing to insufficient astrometric accuracy.  The sloping dotted line
indicates expected improvement in the measurement of the proper motion of \sgrab
within $\approx5$~years.

\end{document}